# Towards optimization of photonic-crystal surface-emitting lasers via quantum annealing

**TAKUYA INOUE,**[1,*] **YUYA SEKI,**[2] **SHU TANAKA,**[3] **NOZOMU TOGAWA,**[4] **KENJI ISHIZAKI,**[1] **AND SUSUMU NODA**[1]

[1]*Photonics and Electronics Science and Engineering Center, Kyoto University, Kyoto-daigaku-katsura, Nishikyo-ku, Kyoto 615-8510, Japan*

[2]*Graduate School of Science and Technology, Keio University, 3-14-1 Hiyoshi, Kohoku-ku, Yokohama, Kanagawa 223-8522, Japan*

[3]*Faculty of Science and Technology, Keio University, 3-14-1 Hiyoshi, Kohoku-ku, Yokohama, Kanagawa 223-8522, Japan*

[4]*Faculty of Science and Engineering, Waseda University, 3-4-1 Ookubo, Shinjuku-ku, Tokyo 169-8555, Japan*

*\*t_inoue@qoe.kuee.kyoto-u.ac.jp*

**Abstract:** Photonic-crystal surface-emitting lasers (PCSELs), which utilize a two-dimensional (2D) optical resonance inside a photonic crystal for lasing, feature various outstanding functionalities such as single-mode high-power operation and arbitrary control of beam polarizations. Although most of the previous designs of PCSELs employ spatially uniform photonic crystals, it is expected that lasing performance can be further improved if it becomes possible to optimize the spatial distribution of photonic crystals. In this paper, we investigate the structural optimization of PCSELs via quantum annealing towards high-power, narrow-beam-divergence operation with linear polarization. The optimization of PCSELs is performed by the iteration of the following three steps: (1) time-dependent 3D coupled-wave analysis of lasing performance, (2) formulation of the lasing performance via a factorization machine, and (3) selection of optimal solution(s) via quantum annealing. By using this approach, we successfully discover an advanced PCSEL with a non-uniform spatial distribution of the band-edge frequency and injection current, which simultaneously enables higher output power, a narrower divergence angle, and a higher linear polarization ratio than conventional uniform PCSELs. Our results potentially indicate the universal applicability of quantum annealing, which has been mainly applied to specific types of discrete optimization problems so far, for various physics and engineering problems in the field of smart manufacturing.

## 1. Introduction

The demand for high-power, high-beam-quality semiconductor lasers is rapidly increasing nowadays for a variety of applications such as light detection and ranging (LiDAR) in smart mobility and high-precision laser processing in smart manufacturing [1–4]. Photonic-crystal surface-emitting lasers (PCSELs) [5–13] are leading candidates to answer this demand, because a two-dimensional (2D) optical resonance at a singularity point (Γ point, etc.) of their photonic band enables coherent lasing operation in a much larger area than conventional semiconductor lasers such as edge-emitting lasers [14] or vertical-cavity surface-emitting lasers [15]. Towards the improvement of output power and beam quality of PCSELs, we previously proposed a double-lattice photonic crystal, in which two lattice-point groups are shifted in the $x$ and $y$ directions by one quarter of the lattice constant [9]. In the double-lattice photonic crystal, optical feedback is weakened by destructive interference, whereby the differences in optical losses between the fundamental mode and the higher-order modes are drastically increased. Using this lattice-point design, we experimentally demonstrated 10-W-to-20-W-class single-mode lasing with PCSELs with a diameter of as large as 400–500 μm [11,12], and we theoretically established the general recipe to realize ultra-large-area (3 mm–10 mm) PCSELs with 100-W-to-1-kW single-mode operation [13].

While PCSELs have achieved various unprecedented performances as described above, the devices demonstrated so far have utilized only a small portion of their potential degrees of freedom in their design; for example, most of the previously demonstrated PCSELs employed a spatially uniform photonic-crystal cavity in the entire device, even though they can potentially employ an arbitrary spatial distribution of varying lattice constants or varying hole shapes. To design such photonic crystals with non-uniform spatial distributions, we have to consider a much larger number of design parameters than in conventional uniform PCSELs, and this number increases exponentially as the device size becomes larger. Therefore, such structural optimization of PCSELs has been a challenging task and has yet to be achieved.

In this paper, we investigate the structural optimization of PCSELs using quantum annealing [16,17] towards high-power, narrow-divergence operation with linear polarization. Quantum annealing is attracting increasing attention in various fields [18–24] owing to its superior capability of solving specific types of discrete optimization problems described with quadratic unconstrained binary optimization (QUBO). To apply quantum annealing for the optimization of PCSELs, we first calculate various lasing performances of PCSELs (output power, beam divergence, polarization, etc.) by time-dependent 3D coupled-wave analysis [25], and we define a figure of merit (FOM) to optimize. Second, we formulate the FOM in the framework of QUBO by a factorization machine [24,26,27]. Third, we search the candidates of the best structures which maximize the FOM using quantum annealing. Repeating these three steps, we discover an optimized device structure with non-uniform spatial distributions of the band-edge frequency and injection current, which simultaneously enables higher output power, a narrower divergence angle, and a higher linear polarization ratio than conventional PCSELs with uniform photonic crystals.

## 2. Methods

The simulation model of the PCSEL device is shown in Fig. 1. Figure 1(a) shows the cross section of the device, where active layers (3-pair InGaAs/AlGaAs quantum wells) and a photonic-crystal layer are placed between p-cladding and n-cladding layers for current injection.

The lasing wavelength of the PCSEL is set as ~940 nm. In this study, we assume the outer diameter of the p-side electrode(s) $L$=1 mm. It should be noted that we can control the current injection distribution of the device by changing the design of the p-side electrode(s) such as into multiple ring-shape electrodes. As for the photonic-crystal structure, we employ a double-lattice photonic crystal with elliptic and circular holes shown in Fig. 1(b) [13]. In this structure, destructive interference between 180° diffraction and 90° diffraction $\kappa_{1D} + \kappa_{2D-}$ determines the complex band structure of the lasing band (band A) near the Γ point as well as the threshold margin between the fundamental mode and higher-order modes in the finite-size cavity [13]. The complex value of $\kappa_{1D} + \kappa_{2D-}$ can be arbitrarily controlled through fine adjustment of the hole distance ($d$) and hole-size balance ($x$) as shown in Fig. 1(c), which enables stable single-mode lasing operation according to the device size.

In the following optimization, we change the following structural parameters: (i) real and imaginary parts of $\kappa_{1D} + \kappa_{2D-}$ (which corresponds to $d$ and $x$ of the double-lattice structure), (ii) spatial distribution of the band-edge frequency $f(\mathbf{r})$ (which corresponds to the spatial distribution of the lattice constant of the photonic crystal), and (iii) spatial distribution of the injection current density $J(\mathbf{r})$ (which can be controlled with the design of the p-side electrodes). Although we can consider arbitrary spatial functions for $f(\mathbf{r})$ and $J(\mathbf{r})$ in principle, here we focus on a concentric frequency distribution and a Gaussian current distribution as shown in Figs. 1(d) and 1(e), which are described with the following equations.

$$f(r) = \sum_{m=0}^{N} f_m \cos\left(\frac{m\pi}{L} r\right), \tag{1}$$

$$J(r) = J_0 \exp\left(-r^2/r_{\text{gauss}}^2\right) \times \frac{1}{1+\exp\left[(r-L/2)/r_j\right]}. \tag{2}$$

Here, $r$ denotes the distance from the center of the current injection region, $f_m$ is the Fourier expansion coefficient of the band-edge frequency distributions, $N$ is the maximum order of the

Fourier expansion ($N=7$ in this study), $r_{gauss}$ is the Gaussian width, and $r_j$ is the lateral current spread outside the electrode ($r_j=25$ μm in this study). The second term of the right hand side of Eq. (2) signifies the decay of the current density outside the electrode. The total amount of the injection current in the entire device is fixed to 16 A. It should be noted that all design parameters [Re($\kappa_{1D}+\kappa_{2D-}$), Im($\kappa_{1D}+\kappa_{2D-}$), $f_0, f_1, …, f_7, r_{gauss}$] can take continuous real values while quantum annealing can only deal with binary values (0 or 1). Therefore, we discretize each parameter into 16 values by representing it with a 4-bit binary number [27]. Consequently, we use 44 q-bits in total for the following optimizations.

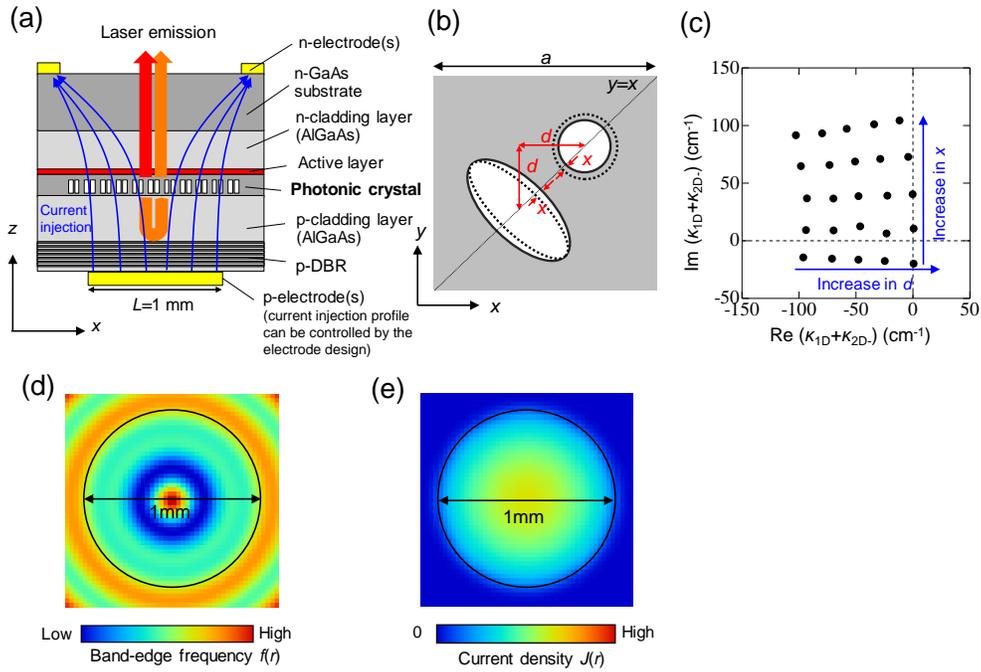

**Fig. 1.** Simulation model of PCSELs. (a) Cross-section of a PCSEL. (b) Schematic of a double-lattice photonic crystal. Hole distance $d$ and hole-size balance $x$ are structural parameters to determine lasing characteristics. (c) Coupling coefficients of the photonic crystal when $d$ and $x$ are changed. (d)(e) Examples of spatial distributions of band-edge frequency $f(r)$ and current density $J(r)$.

The optimization procedure is outlined in Fig. 2. In Step (I), we prepare one uniform structure (initial structure) and 9 random structures, and we perform a comprehensive analysis of the lasing performance for each structure by time-dependent 3D-CWT [25], which considers both the spatial and temporal evolution of the carrier-photon interactions inside PCSELs. The parameters used for the 3D-CWT simulation are summarized in Appendix A. In this study, we aim to improve the following three lasing characteristics: (i) output power at an injection current of 16A ($P$), (ii) beam divergence angles in the $x$ and $y$ directions evaluated with D4σ values ($\theta_x$ and $\theta_y$), and (iii) polarization ratio ($\eta$), which is defined by the ratio between the output power along the main polarization axis and that along the perpendicular axis. As for the polarization ratio, the beam emitted from typical PCSELs is not perfectly linearly polarized; although the emission from band A in an infinite double-lattice photonic crystal is linearly polarized along the line of $y=-x$, the emission from a finite-size cavity contains perpendicular polarizations (along the line of $y=x$). This is because the electric field distribution near the edge of the current injection region is different from that at the center.

In general, there are tradeoffs among the improvement of the above three characteristics; for example, current injection with a smaller Gaussian width ($r_{gauss}$) leads to a decrease of the threshold current and an increase of the output power, while the beam divergence angle increases due to the smaller aperture and the polarization ratio decreases due to the stronger finite-size effect. To simultaneously achieve high output power, a narrow divergence angle, and a higher polarization ratio, we define the following dimensionless FOM ($Q$) for the optimization;

$$Q = \frac{P \times \eta}{\theta_x \theta_y} \times 10^{-3} \text{ W}^{-1} \text{deg}^2. \tag{3}$$

After calculating the FOM in Eq. (3) for the ten initial structures, we formulate the above FOM in the framework of QUBO [Step (II)], which is described as follows;

$$Q = w_0 + \sum_{i=1}^{44} w_i q_i + \sum_{i=1}^{44} \sum_{j>i}^{44} w_{ij} q_i q_j \ . \qquad (4)$$

Here, $q_i$ denotes each binary parameter (0 or 1) and $w_0$, $w_i$, $w_{ij}$ are fitting coefficients. The values of these coefficients are initially given by random values, and then are renewed according to the algorithm of the factorization machine [26,27] so that the error of $Q$ is minimized. In this study, we employ the factorization size of 10 and use the algorithm of stochastic gradient descent for the parameter renewal.

In Step (III), using the formulation described in Eq. (4), we perform quantum annealing using the D-Wave Advantage[TM] quantum annealer [28] and find the parameter set(s) $\{q_i\}$ to maximize $Q$ (or minimize $-Q$). Since this annealing process is stochastic, we repeat the annealing processes 1000 times and choose five solutions in order of largest $Q$. Then, we add these parameter sets to the current sample set, and we iterate steps (I)-(III) until the FOM value converges.

It should be noted that in the optimization procedure shown in Fig. 2, Step (I) (calculation of $Q$) takes a far longer time than the other steps, and thus the number of $Q$-calculations should be decreased as much as possible. To evaluate the usefulness of the quantum annealing, we also perform the optimization with several classical algorithms (specifically, a genetic algorithm [29] and particle swarm optimization (PSO) [30]), and we compare the number of $Q$-calculations required to obtain the optimized structure.

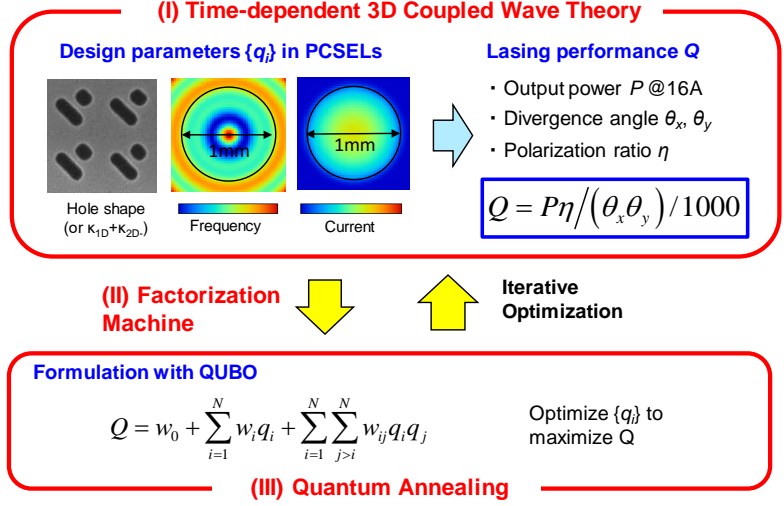

**Fig. 2.** Procedure of structural optimization of PCSELs. The optimization is performed by iteration of the following three steps: (I) comprehensive analysis of a figure of merit (FOM) via time-dependent 3D coupled-wave analysis, (II) formulation of the FOM in the framework of quadratic unconstrained binary optimization (QUBO) via a factorization machine, and (III) selection of optimal solution(s) via quantum annealing.

## 3. Results

The evolution of the best FOM ($Q$) for 1-mm PCSELs during the optimization process is shown in Fig. 3(a). Here, the results for quantum annealing (red) and those for other classical algorithms [Genetic algorithm (black), PSO with 10 particles (blue), and PSO with 20 particles (yellow green)] are compared. In the case of quantum annealing, we obtain an optimized structure with a FOM of 11.64, which is over 3.8 times larger than the initial uniform structure ($Q_{init}$=3.08). We repeated the optimization process using a different set of initial random structures, and we obtained almost the same value of FOM (11.60), which confirmed the robustness of the process. In contrast, when we applied the other classical algorithms, we obtained lower FOMs, which indicate convergence to local maxima. In addition, the number of $Q$-calculations required to obtain a high FOM ($Q>3Q_{init}$, for example) is much smaller for quantum annealing than for the other classical algorithms. Although the optimization speeds of these algorithms might depend on the selection of the initial parameters, these results indicate the potential usability of quantum annealing in the structural optimization of photonic devices.

Figure 3(b) shows a comparison between the initial structure and the optimized structure via quantum annealing. The real and imaginary parts of $\kappa_{1D}+\kappa_{2D-}$ are automatically adjusted after optimization, which correspond to a change of +0.7 nm in $d$ and +0.8 nm in $x$ in the double-lattice structure. The above change of $\kappa_{1D}+\kappa_{2D-}$ leads to an increase of the radiation constant (for an infinite structure) from 11.0 cm$^{-1}$ to 35.7 cm$^{-1}$, which contributes to the increase of the output power along the main polarization axis (along $y=-x$). The band-edge frequency distribution of the optimized structure exhibits a non-uniform pattern, where the frequency change near the boundary of the 1-mm-diameter electrode is especially large. Interestingly, such an abrupt change of the band-edge frequency at the boundary of the electrode is similar to the pop-down structure we proposed in our previous paper [31], which compensates for the refractive-index difference caused by the carrier-density distributions inside and outside the current injection region. It should be also noted that the frequency distribution in the outermost region [outside the 1.2 mm-diameter circle shown in Fig. 3(b)] does not affect the lasing performance owing to the negligibly small current injection there. The current distribution of the optimized structure is a Gaussian profile with a width of $2r_{gauss}$=0.73 mm.

Figures 3(c) to 3(e) shows the current-light-output characteristics, far-field beam profiles, and polarization-resolved near-field patterns of the initial and optimized structures via quantum annealing. In Fig. 3(c), the slope efficiency of the optimized structure is higher than that of the initial structure, which is due to the increase of the radiation constant of the double-lattice photonic crystal as described above. In the far-field beam profile shown in Fig. 3(d), the side-lobe emission intensity of the optimized structure is one order of magnitude lower than that of the initial structure, which contributes to the smaller beam divergence angle evaluated with D4σ. The reduction of the side-lobe intensity is ascribed to the Gaussian current injection profile, which realizes Gaussian confinement of the lasing mode. Further reduction of the side-lobe intensity (below -30 dB, for example) is expected by directly including the side-lobe intensity ratio in the definition of the FOM. As for the polarization-resolved near-field profiles [Fig. 3(e)], the emission from the initial structure contains not only the polarization along $y=-x$ but also that along $y=x$, where the latter polarized light is mainly emitted from the edge of the device as explained before. In contrast, the emission from the optimized structure shows a much larger power contrast between the two polarizations, owing to the combination of the larger radiation constant and the Gaussian current injection profile.

Table I shows a summary of the lasing performance of the initial structure and the optimized structure. The optimized structure simultaneously enables higher output power, a narrower divergence angle, and a higher linear polarization ratio than the initial, conventionally uniform structure.

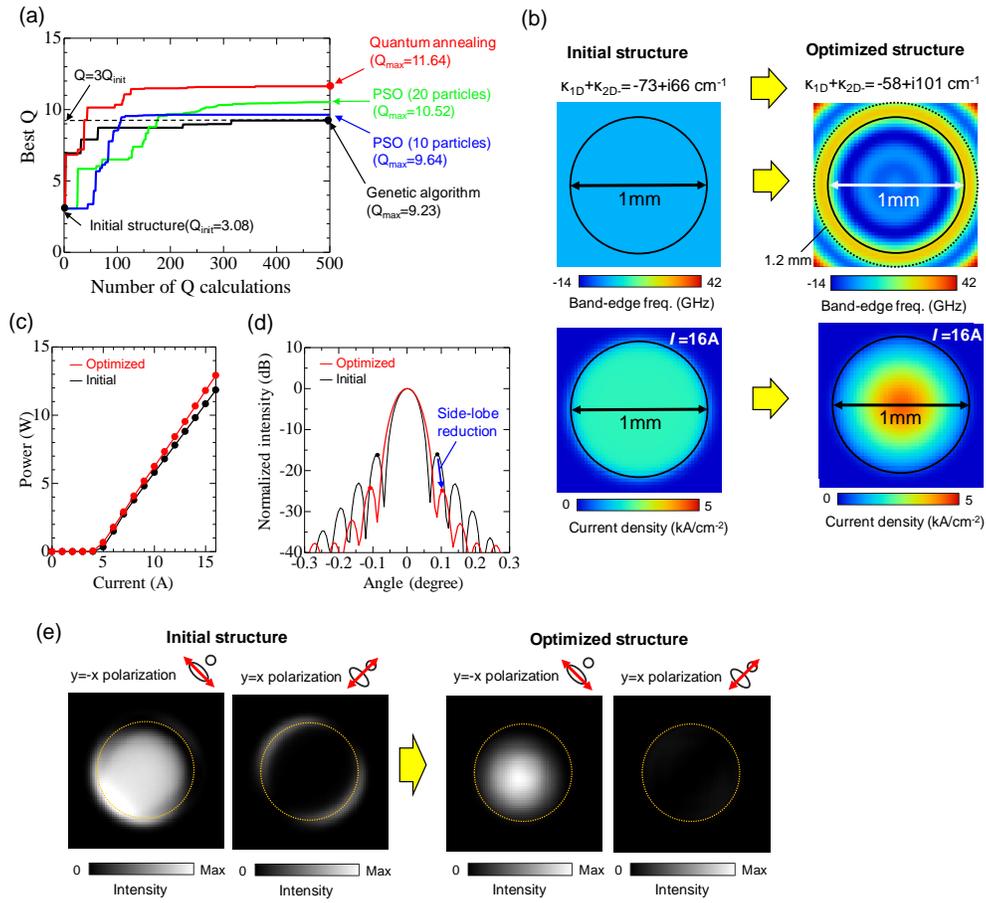

**Fig.3.** (a) Evolution of the best FOM ($Q$) for 1-mm PCSELs during optimization by various optimization methods. (b) Comparison of the initial structure and the optimized structure via quantum annealing. (c) Current-light-output characteristics of the initial structure and the optimized structure. (d) Far-field beam profile of the initial structure and the optimized structure. (e) Polarization-resolved near-field pattern of the initial structure and the optimized structure.

**Table 1. Calculated lasing performance before and after quantum optimization**

|  | Initial structure | Optimized structure |
| --- | --- | --- |
| Output power at 16A $P$ (W) | 11.8 | 13.0 |
| Divergence angle $\theta_x$/$\theta_y$ (degree) | 0.154/0.154 | 0.124/0.124 |
| Polarization ratio | 6.2 | 13.5 |

## 4. Conclusion

We have investigated the optimization of the device structure of PCSELs using quantum annealing, towards the realization of high-power, narrow-divergence operation with linear polarization. We have established a method of iterative optimization combining time-dependent 3D-CWT, a factorization machine, and quantum annealing, and we successfully discovered an optimized structure that can simultaneously realize higher output power, a narrower divergence angle, and a higher linear polarization ratio than conventional, uniform PCSELs. Our quantum-assisted optimization method is potentially advantageous compared to classical methods in that it can reach the (nearly) optimized solution over a smaller number of iterations. It should be noted that the optimization method we established in this study can be also applied to 100-W-to-1-kW-class PCSELs with larger diameters (>3 mm) [13], which will impact various laser-related industries including laser sensing for smart mobility, laser manufacturing, and laser medicine. More generally, our results have potentially indicated the universal applicability of quantum annealing, which has been mainly applied to specific types of discrete optimization problems so far, for various physics and engineering problems in the field of smart manufacturing [32]. We hope that our results will facilitate the use of quantum technologies for the development of a smart society in the future.

# Appendix A: Parameters used for 3D-CWT simulations

Structural parameters of the designed PCSEL and other parameters used for time-dependent 3D-CWT [25] are summarized in Table 2 and Table 3, respectively.

**Table 2. Structural parameters of the PCSEL**

| Layer | Thickness (nm) | Refractive index |
|---|---|---|
| n-cladding (AlGaAs) | 1000 | 3.38 |
| AlGaAs | 300 | 3.45 |
| Active (InGaAs/AlGaAs) | (10/20)×3 | 3.58/3.45 |
| AlGaAs | 25 | 3.27 |
| GaAs | 90 | 3.55 |
| PC | 160 | $n_{pc}$ |
| p-cladding (AlGaAs) | 980 | 3.32 |
| DBR (AlGaAs/AlGaAs) | (68/78)×14 | 3.47/3.01 |
| Contact (GaAs) | 300 | 3.55 |

**Table 3. Parameters used for time-dependent 3D-CWT simulations**

| Symbol | Parameter | Value |
|---|---|---|
| $a$ | Lattice constant | 276 nm |
| $n_g$ | Group refractive index | 3.451 |
| $n_{eff}$ | Effective refractive index | 3.43 |
| $\alpha_0$ | Internal material loss | 3 cm$^{-1}$ |
| $d_{active}$ | Total thickness of quantum wells | 30 nm |
| $\Gamma_{active}$ | Optical confinement factor | 0.06 |
| $\tau_C$ | Carrier lifetime | 2 ns |
| $D$ | Diffusion constant | 100 cm$^2$s$^{-1}$ |
| $dn/dN$ | Carrier density coefficient of refractive index | $-7.6\times10^{-21}$ cm$^3$ |
| $\beta$ | Spontaneous emission factor | $1.0\times10^{-4}$ |